\documentclass{article} %

\makeatletter
\RequirePackage{tgpagella} %
\RequirePackage{mathpazo}  %
\RequirePackage{fontawesome5}
\RequirePackage{inconsolata} %

\usepackage{eso-pic} %
\RequirePackage{fancyhdr}
\RequirePackage{xcolor}
\RequirePackage{natbib}
\RequirePackage[a4paper,margin=2.5cm,headheight=52pt]{geometry}

\RequirePackage[style=iso]{datetime2}
\RequirePackage{tikz}
\RequirePackage{titling}

\RequirePackage{microtype}
\RequirePackage{setspace}
\RequirePackage{hyperref}
\RequirePackage{url}
\definecolor{darkblue}{rgb}{0, 0, 0.5}
\hypersetup{colorlinks=true, citecolor=darkblue, linkcolor=darkblue, urlcolor=darkblue}

\setcitestyle{authoryear,round,citesep={;},aysep={,},yysep={;}}

\def\addcontentsline#1#2#3{}

\def\maketitle{\par
\begingroup
   \def\thefootnote{\fnsymbol{footnote}}
   \def\@makefnmark{\hbox to 0pt{$^{\@thefnmark}$\hss}} %
   \long\def\@makefntext##1{\parindent 1em\noindent
                            \hbox to1.8em{\hss $\m@th ^{\@thefnmark}$}##1}
   \@maketitle \@thanks
\endgroup
\setcounter{footnote}{0}
\let\maketitle\relax \let\@maketitle\relax
\gdef\@thanks{}\gdef\@author{}\gdef\@title{}\let\thanks\relax}

\usepackage{fancyhdr}
\renewcommand{\headrulewidth}{1pt}
\renewcommand{\headrule}{{\color[HTML]{4D4D4D}\hrule\@height\headrulewidth\@width\headwidth\vskip-\headrulewidth}}

\def\twodigits#1{\ifnum#1<10 0\fi\the#1}
\fancypagestyle{firstpage}{
  \lhead{%
    \raisebox{-0.5\height}{\includegraphics[height=42pt]{aslp-logo.pdf}}%
    \hspace{10pt}%
    \begingroup
      \setlength{\fboxsep}{2pt}%
      \raisebox{-0.5\height}{\colorbox[HTML]{123B5D}{%
        \includegraphics[height=22pt]{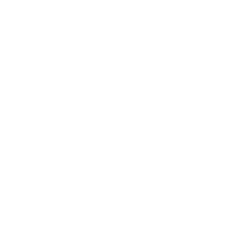}}}%
    \endgroup
  }}

  \renewcommand{\headrule}{{\color[HTML]{4D4D4D}\hrule\@height\headrulewidth\@width\headwidth\vskip-\headrulewidth}}

\def\huggingface{\raisebox{-1.5pt}{\includegraphics[height=1.15em]{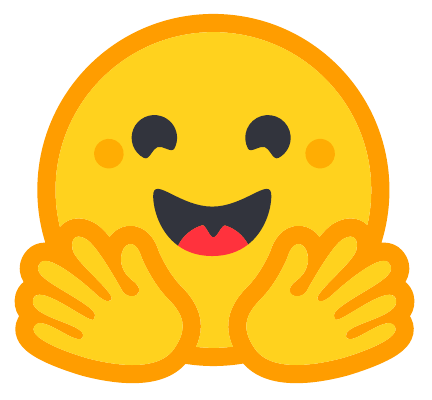}}}
\def\github{\raisebox{-1.5pt}{\includegraphics[height=1.15em]{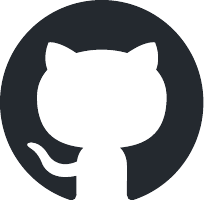}}}

\newcommand{\ghlink}{https://github.com/ASLP-lab/DRAMA-Eval}
\newcommand{\hflink}{https://huggingface.co/datasets/ASLP-lab/DRAMA-Eval}
\newcommand{\resultslink}{https://huggingface.co/datasets/ASLP-lab/DRAMA-Eval-Results}
\newcommand{\demolink}{https://ASLP-lab.github.io/DRAMA-Eval-Demo}

\newcommand{\demosparkles}{%
  \begin{tikzpicture}[x=1em,y=1em,baseline=-0.15em]
    \fill[orange!85!yellow] (0.55,1.0)--(0.68,0.62)--(1.0,0.5)--(0.68,0.38)--(0.55,0.0)--(0.42,0.38)--(0.1,0.5)--(0.42,0.62)--cycle;
    \fill[orange!85!yellow] (0.1,1.0)--(0.15,0.86)--(0.3,0.81)--(0.15,0.76)--(0.1,0.62)--(0.05,0.76)--(-0.1,0.81)--(0.05,0.86)--cycle;
    \fill[orange!85!yellow] (0.96,0.27)--(1.0,0.16)--(1.11,0.12)--(1.0,0.08)--(0.96,-0.03)--(0.92,0.08)--(0.81,0.12)--(0.92,0.16)--cycle;
  \end{tikzpicture}%
}

\def\@maketitle{\vbox{\hsize\textwidth
 {\centering {\Large\bf \@title\par}}
\vspace{-0.3cm}
 \begin{quote}\rule{\z@}{6pt} {\centering\par{\@author}\par}\end{quote}%
\vskip 0.3in minus 0.1in
}%
\begin{center}
\vspace{-1cm}
\begin{tabular}{rl}
\github & \url{\ghlink}\\
\huggingface & \url{\hflink}\\
\href{\resultslink}{\textcolor[HTML]{E56B3F}{\faRocket}} & \url{\resultslink}\\
\href{\demolink}{\demosparkles} & \url{\demolink}\\
\end{tabular}
\end{center}
\thispagestyle{firstpage}}

\renewenvironment{abstract}{\vskip.075in\centerline{\large\bf
Abstract}\vspace{0.5ex}\begin{quote}}{\par\end{quote}\vskip 1ex}

\def\section{\@startsection {section}{1}{\z@}{-2.0ex plus
    -0.5ex minus -.2ex}{1.5ex plus 0.3ex
minus0.2ex}{\large\bf\raggedright}}

\def\subsection{\@startsection{subsection}{2}{\z@}{-1.8ex plus
-0.5ex minus -.2ex}{0.8ex plus .2ex}{\normalsize\bf\raggedright}}
\def\subsubsection{\@startsection{subsubsection}{3}{\z@}{-1.5ex
plus      -0.5ex minus -.2ex}{0.5ex plus
.2ex}{\normalsize\bf\raggedright}}
\def\paragraph{\@startsection{paragraph}{4}{\z@}{1.5ex plus
0.5ex minus .2ex}{-1em}{\normalsize\bf}}
\def\subparagraph{\@startsection{subparagraph}{5}{\z@}{1.5ex plus
  0.5ex minus .2ex}{-1em}{\normalsize}}

\skip\footins 9pt plus 4pt minus 2pt
\def\footnoterule{\kern-3pt \hrule width 12pc \kern 2.6pt }
\leftmargini\leftmargin \leftmarginii 2em
\def\@listi{\leftmargin\leftmargini}
\def\@listii{\leftmargin\leftmarginii
   \labelwidth\leftmarginii\advance\labelwidth-\labelsep
   \topsep 2pt plus 1pt minus 0.5pt
   \parsep 1pt plus 0.5pt minus 0.5pt
   \itemsep \parsep}
\def\@listiii{\leftmargin\leftmarginiii
    \labelwidth\leftmarginiii\advance\labelwidth-\labelsep
    \topsep 1pt plus 0.5pt minus 0.5pt
    \parsep \z@ \partopsep 0.5pt plus 0pt minus 0.5pt
    \itemsep \topsep}
\def\@listiv{\leftmargin\leftmarginiv
     \labelwidth\leftmarginiv\advance\labelwidth-\labelsep}
\def\@listv{\leftmargin\leftmarginv
     \labelwidth\leftmarginv\advance\labelwidth-\labelsep}
\def\@listvi{\leftmargin\leftmarginvi
     \labelwidth\leftmarginvi\advance\labelwidth-\labelsep}

\belowdisplayskip \abovedisplayskip
\def\normalsize{\@setsize\normalsize{11pt}\xpt\@xpt}
\def\small{\@setsize\small{10pt}\ixpt\@ixpt}
\def\footnotesize{\@setsize\footnotesize{10pt}\ixpt\@ixpt}
\def\scriptsize{\@setsize\scriptsize{8pt}\viipt\@viipt}
\def\tiny{\@setsize\tiny{7pt}\vipt\@vipt}
\def\large{\@setsize\large{14pt}\xiipt\@xiipt}
\def\Large{\@setsize\Large{16pt}\xivpt\@xivpt}
\def\LARGE{\@setsize\LARGE{20pt}\xviipt\@xviipt}
\def\huge{\@setsize\huge{23pt}\xxpt\@xxpt}
\def\Huge{\@setsize\Huge{28pt}\xxvpt\@xxvpt}

\makeatother

\usepackage{booktabs}
\usepackage{graphicx}
\usepackage{enumitem}

\usepackage{CJKutf8}

\usepackage{microtype}
\usepackage{amsmath}
\usepackage{amsfonts}
\usepackage{colortbl}
\usepackage[utf8]{inputenc}
\usepackage[T1]{fontenc}
\definecolor{lightgray}{rgb}{0.9,0.9,0.9}
\usepackage{caption}
\usepackage{xcolor}
\usepackage{setspace}
\usepackage{url}
\usepackage{tabularx}
\usepackage{xltabular}
\usepackage{tikz}
\usepackage{makecell}
\usepackage[skins]{tcolorbox}
\usepackage{xurl}

\usepackage{fvextra}
\tcbuselibrary{breakable}
\begin{document}
\begin{figure}[htbp]
    \centering
    \includegraphics[width=0.2\textwidth]{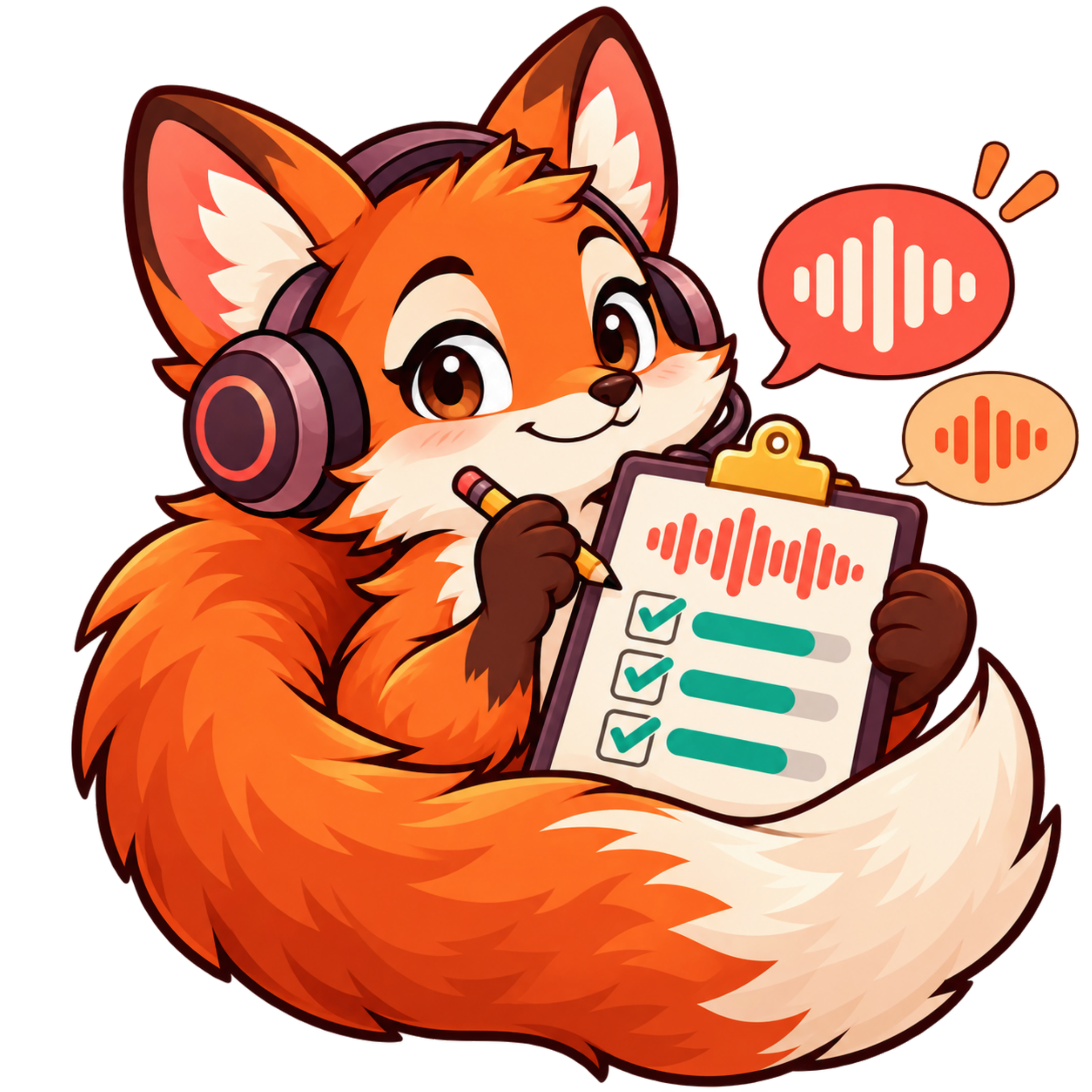}
    \label{fig:logo}
\end{figure}
\title{DRAMA-Eval: A Benchmark for Context-Aware, Multi-Turn,
Multi-Speaker Speech Generation}

\author{
\textbf{Jingbin Hu$^{1,*}$},
\textbf{Zihang Yang$^{1,*}$},
\textbf{Xiaoge Peng$^{2,*}$},
\textbf{Haoyu Zhang$^{1,*}$},
\textbf{Ziyu Zhang$^1$}, \\
\textbf{Qirui Zhan$^1$},
\textbf{Huakang Chen$^1$},
\textbf{Zhao Guo$^1$},
\textbf{Wenjie Tian$^1$},
\textbf{Kangxiang Xia$^1$}, \\
\textbf{Wenkang Hou$^1$},
\textbf{Ruonan You$^1$},
\textbf{Zihan Liu$^1$},
\textbf{Chenyi Liao$^1$},
\textbf{Hongyu Lyu$^1$},
\textbf{Zihao Bai$^1$},
\textbf{Jinrui Liang$^1$},
\textbf{Shengming Lin$^1$},
\textbf{Yunxiang Chen$^2$},
\textbf{Houdun Liu$^2$},
\textbf{Su Feng$^2$},
\textbf{Bengu Wu$^3$},
\textbf{Liumeng Xue$^{4,\dagger}$},
\textbf{Lei Xie$^{1,\dagger}$}
\\[0.5em]
$^1$Audio, Speech and Language Processing Group (ASLP@NPU),\\
\mbox{School of Computer Science, Northwestern Polytechnical University}\\
$^2$Shenzhen Pimei Technology Co., Ltd.\\
$^3$Yutu Zhineng\\
$^4$Audio, Speech and Language Processing Group (ASLP@NJU),\\
School of Intelligence Science and Technology, Nanjing University
}

\maketitle

\begingroup
\renewcommand{\thefootnote}{}
\footnotetext{%
  $^{*}$ Equal contribution (joint first authors).
  $^{\dagger}$ Corresponding authors.
}
\endgroup

\begin{abstract}

Recent advances in speech generation are shifting toward context-aware, multi-turn, and multi-speaker synthesis. While these capabilities are in high demand for applications such as audiobook and audio drama production, evaluation has not kept pace with these advances. Beyond traditional zero-shot Text-to-Speech (TTS), modern systems in this domain primarily fall into three categories: instruction-controllable generation models, multi-turn dialogue models, and the audio branches of audio-visual generation models. However, existing benchmarks evaluate these paradigms in isolation, failing to provide a unified assessment of context modeling, instruction following, and multi-speaker generation. Meanwhile, recent progress in audio understanding models has made it feasible to evaluate such complex scenarios using LLM-as-a-Judge approaches. To bridge this gap, we introduce \textbf{DRAMA-Eval}, a comprehensive benchmark for evaluating such systems. DRAMA-Eval utilizes structured script contexts and offers a holistic evaluation framework covering objective, local, and global aspects of generation quality, enabling the assessment of acoustic fidelity, sentence-level execution, and long-range coherence. We benchmark leading open-source and commercial models on a human-verified dataset. The results reveal critical weaknesses in long-range speaker consistency and implicit instruction following that require deep contextual understanding. To facilitate reproducible research, we release a 12-hour bilingual (Chinese-English) test set of manually corrected real-world scripts, accompanied by ground-truth audio, evaluation code, and benchmark results.
\end{abstract}

\section{Introduction}
Driven by applications such as audiobook and audio drama production, speech generation is rapidly shifting toward script-driven synthesis. This shift calls for holistic evaluation spanning local instruction execution to long-range scene coherence.

Over the past few years, speech synthesis has evolved through several key paradigms. Early zero-shot Text-to-Speech (TTS) models~\cite{valle,cosyvoice,cosyvoice2,cosyvoice3,sparktts,llasa,indextts,indextts2,voxcpm,voxcpm2,moss-tts,f5-tts,maskgct,qwentts,omnivoice, dots,fireredtts3} primarily focused on cloning target voice identities from short reference audio clips. Subsequent work extended this capability to multi-speaker dialogue synthesis~\cite{moss-ttsd,soulxpodcast,vibevoice,fireredtts2}, prioritizing natural turn-taking dynamics in conversational settings. To enhance controllability and expressiveness, instruction-following models~\cite{voxcpm2,qwentts,fireredtts3,moss-vd,hu2026voicesculptor} introduced explicit textual prompts for emotion and style steering. Concurrently, voice control models~\cite{voxcpm2} enabled the decoupling of vocal identity from acoustic style, preserving the original timbre while dynamically modifying emotional tone or delivery. Most recently, Context-Aware TTS models~\cite{cot-tts,cot-tts-up,harness-tts, Interactive_tts, han2026fish} have emerged, leveraging surrounding dialogue history as implicit contextual cues to produce scene-adaptive speech.

In parallel, the audio branches of audio-visual generation models~\cite{mova,ltx,minimaxh3_2025}, further reflect the trend toward context-aware, multi-turn, multi-speaker speech generation.

Despite these technological advances, a fundamental disconnect exists between existing model capabilities and evaluation methodologies. On the one hand, instruction-following models accept rich character and scene descriptions, yet they target isolated utterances or short speech clips. On the other hand, current open-source multi-turn dialogue models preserve speaker identity across turns but primarily condition on plain transcripts and speaker IDs rather than structured, script-level context. In the meanwhile, the evaluation of the audio branches of audio-visual generation models remains limited. Current assessments predominantly focus on surface-level acoustic properties, such as audio quality, intelligibility, and basic expressiveness~\cite{audiobox}, while overlooking higher-level capabilities essential for complex multi-turn scenarios, including contextual instruction following, natural multi-speaker turn-taking, and speaker identity consistency over extended durations.

To address these limitations, we propose \textbf{DRAMA-Eval}, a benchmark designed to systematically evaluate multi-turn, multi-speaker speech generation with structured script context~\cite{from_script}. Unlike prior benchmarks that isolate instruction following, utterance-level expressiveness, multi-turn generation, or speaker similarity, DRAMA-Eval provides a unified evaluation protocol that jointly measures objective fidelity, local execution quality, and global consistency. This design enables comprehensive assessment of whether models can understand structured context, follow implicit and explicit instructions, maintain speaker identity across extended interactions, and preserve coherence throughout an entire scene. Furthermore, DRAMA-Eval incorporates A and B preference testing against ground-truth audio using LLM-as-a-Judge approaches~\cite{emergent}, offering a more reliable evaluation protocol for complex dramatic generation. In this way, our benchmark fills a key evaluation gap and supports progress toward production-grade narrative speech generation.

Our main contributions are summarized as follows:
\begin{itemize}
\item We build and release \textbf{DRAMA-Eval}, including a \textbf{12-hour bilingual (Chinese-English)} test set of manually corrected real-world scripts paired with ground-truth audio, broadly sourced from diverse scenarios such as films and TV dramas, podcasts, interviews and so on, and designed for evaluating context-aware, multi-turn, multi-speaker speech generation.

\item We propose a holistic evaluation protocol spanning \textbf{objective}, \textbf{local}, and \textbf{global} aspects of generation quality. In particular, we adopt \textbf{LLM-as-a-judge} with A and B preference comparisons against ground-truth audio to compute win rates, and further validate the reliability of this protocol through human consistency studies.

\item We systematically benchmark a diverse set of leading \textbf{open-source and commercial} models, providing a comprehensive analysis of current speech generation systems and revealing persistent weaknesses in \textbf{long-range consistency} and \textbf{context-driven implicit instruction following}.

\end{itemize}

\label{sec:intro}

\section{Related Work}

\begin{table*}[t]
\centering
\small
\setlength{\tabcolsep}{5pt}
\renewcommand{\arraystretch}{1.15}
\renewcommand{\tabularxcolumn}[1]{m{#1}}

\caption{
Comparison with related benchmarks.
Instr.: instruction-following evaluation;
Identity: cross-utterance speaker consistency;
Scene: complete multi-speaker interaction evaluation.
$\checkmark$ indicates a dedicated evaluation target;
``--'' indicates otherwise.
}
\label{tab:benchmark_comparison}

\begin{tabularx}{\textwidth}{
@{}l c >{\raggedright\arraybackslash}X c c c c@{}
}
\toprule
Benchmark
& Lang.
& Context / conditioning
& \makecell{Evaluation\\unit}
& Instr.
& Identity
& Scene \\
\midrule

\makecell[l]{InstructTTSEval\\
~\cite{instructttseval}}
& EN, ZH
& Acoustic, style, and role instructions
& Utterance
& $\checkmark$ & -- & -- \\
\addlinespace[3pt]

\makecell[l]{MINT-Bench\\
~\cite{mint-bench}}
& 10
& Attribute and style instructions
& Utterance
& $\checkmark$ & -- & -- \\
\addlinespace[3pt]

\makecell[l]{CEAEval\\
~\cite{CEAEval}}
& ZH
& Narrative context for a target utterance
& Utterance
& -- & -- & -- \\
\addlinespace[3pt]

\makecell[l]{SwanBench-Speech\\
~\cite{SwanBench}}
& EN, ZH
& Long-form texts and speaker-labeled dialogues
& Sequence
& -- & $\checkmark$ & $\checkmark$ \\

\midrule

\textbf{DRAMA-Eval (ours)}
& EN, ZH
& \textbf{Scene context, character profiles,
and turn-level instructions}
& Scene
& $\checkmark$ & $\checkmark$ & $\checkmark$ \\

\bottomrule
\end{tabularx}
\end{table*}

\subsection{History of Speech Generation Evaluation}

Speech generation evaluation has traditionally emphasized acoustic quality, intelligibility, and speaker similarity. Recent benchmarks extend this scope to controllability and contextual expressiveness. InstructTTSEval and MINT-Bench~\cite{instructttseval,mint-bench} evaluate natural-language instruction following across acoustic and stylistic attributes, while VStyle~\cite{vstyle} examines voice style adaptation to spoken instructions, including acoustic control, role-play, and implicit empathy. CAST~\cite{knowing-what} evaluates discourse-conditioned prosodic emphasis, and CEAEval~\cite{CEAEval} assesses whether expressive delivery matches the communicative intent implied by narrative context. Speech-DRAME~\cite{speech-drama} further evaluates speech role-play through archetype-oriented and human-grounded realism assessments, incorporating character profiles, scene descriptions, and human-aligned evaluation models. These studies establish that expressive quality depends on both the requested delivery and its contextual appropriateness.

Evaluation has also expanded to long-form speech and multi-turn dialogue. SwanBench-Speech~\cite{SwanBench} covers diverse long-form and dialogue scenarios through acoustic, semantic, and expressive dimensions, including acoustic consistency, prosodic coherence, and expressive hierarchy. MSBench~\cite{xu2026msbench} evaluates one-pass multi-speaker dialogue generation through character adherence, emotional expression, acoustic quality, topic relevance, and dialogue structure. TTSD-eval~\cite{moss-ttsd} examines speaker similarity in multi-speaker dialogue. More directly related to dramatic performance, SceneTTS-Bench~\cite{scenetts-bench} evaluates drama dubbing through cross-turn timbre consistency, emotional expressiveness, and rhythm coherence. Together, these benchmarks demonstrate that long-form evaluation already encompasses consistency, expressive dynamics, and interaction-level requirements.

\subsection{LLM-as-a-Judge for Speech Generation Evaluation}

Large audio-language models have increasingly been adopted as automated judges to assess speech generation beyond conventional objective metrics. Existing benchmarks instantiate this LLM-as-a-judge paradigm with different judge models, evaluation criteria, and judgment protocols.

InstructTTSEval~\cite{instructttseval} employs Gemini 2.5 Pro Preview as the judge to assess instruction following across acoustic-parameter specification, descriptive-style directives, and role-play, focusing on whether generated speech realizes the requested acoustic and expressive characteristics. EmergentTTS-Eval~\cite{emergent} uses Gemini 2.5 Pro as the primary judge to compare candidate speech against a reference system across six categories covering emotion, paralinguistics, and challenging linguistic phenomena. AnyAudio-Judge~\cite{any-audio-judge} instead trains a dedicated audio-language judge based on Qwen3-Omni~\cite{qwen3-omni} to assess fine-grained instruction, audio alignment across speech, sound, music, and mixed audio. Guided by LLM-generated rubrics, the judge verifies individual instruction requirements and aggregates their satisfaction probabilities into an interpretable alignment score. Together, these studies establish model-based judging as a scalable evaluation approach, while illustrating how judge selection and evaluation protocols can be tailored to different aspects of generated audio.

\section{DRAMA-Eval}
\subsection{Task Definition and Problem Formulation}

\begin{figure*}[t]
  \includegraphics[width=\linewidth]{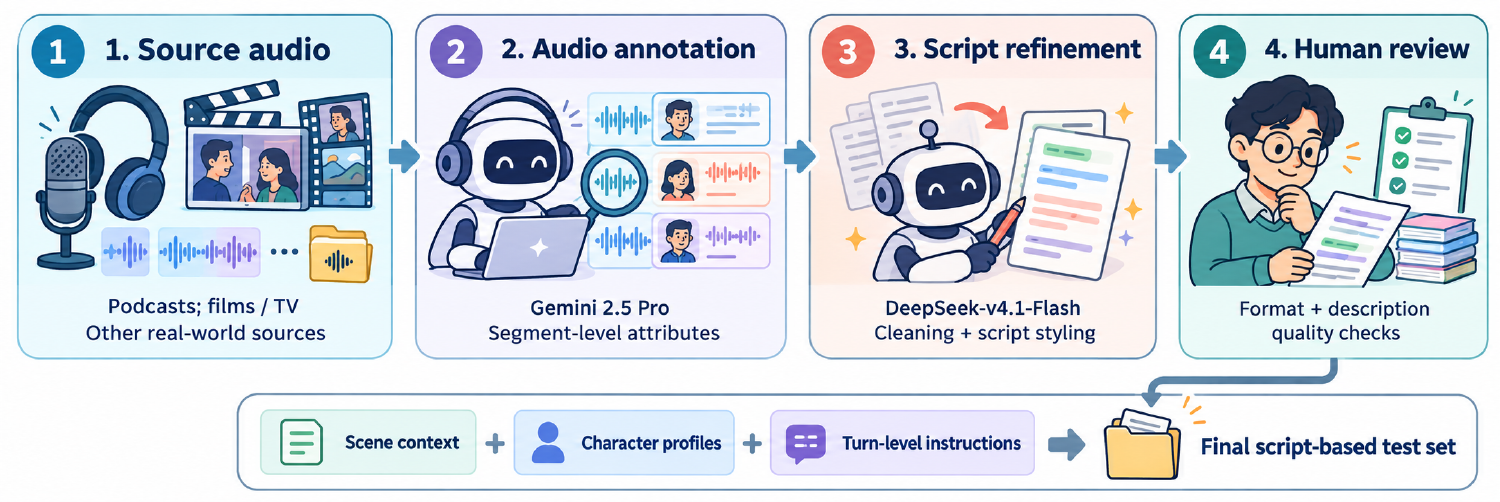}
  \caption{Data construction pipeline of DRAMA-Eval, comprising real-world audio collection, automatic annotation, script refinement, and human review.}
  \label{fig:drama_dataset_pipeline}
\end{figure*}
\begin{figure*}[t]
  \includegraphics[width=0.9\linewidth]{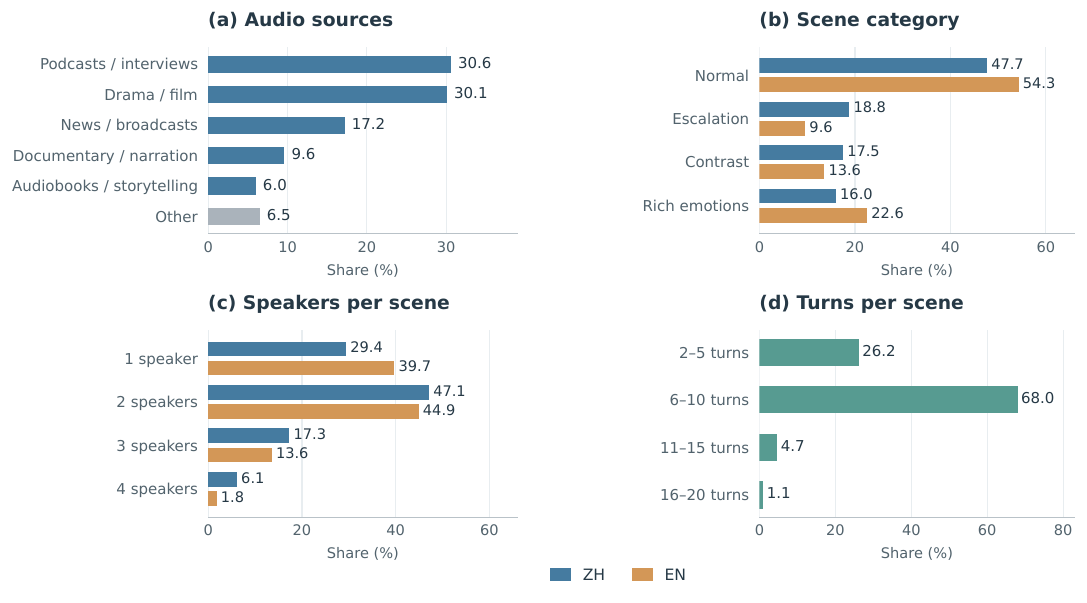}
  \caption{Distributions of audio sources, scene categories, speaker counts, and turn counts in DRAMA-Eval. All values are percentages.}
  \label{fig:dataset_dist}
\end{figure*}
Existing speech generation paradigms emphasize different capabilities, leaving their joint realization in extended, multi-speaker scenes insufficiently examined. Dialogue TTS supports multi-speaker generation, but its ability to execute expressive instructions and deliver context-appropriate performances requires closer evaluation. Instruction-controllable TTS offers fine-grained control over individual utterances, yet utterance-level success does not establish multi-turn speaker consistency or natural transitions between separately synthesized turns. Joint audio-visual generation models provide another route to expressive, immersive scenes, while raising questions about sustained dialogue generation and consistent character voices over longer durations. These complementary strengths and limitations motivate a shared evaluation setting: can a system jointly maintain character identity, realize expressive instructions, and sustain coherent interactions throughout an extended scene?

As summarized in Table~\ref{tab:benchmark_comparison}, DRAMA-Eval judges this question through multi-turn, multi-speaker speech generation conditioned on structured scripts. Its purpose is to expose capability gaps that may remain hidden when instruction following, speaker consistency, and perceptual quality are evaluated separately.

Formally, a structured script is represented as
\[
\mathcal{S}
=
\left(
c,\mathcal{R},
\{(r_t,x_t,u_t)\}_{t=1}^{T}
\right),
\]
where \(c\) specifies the scene context, \(\mathcal{R}\) contains character descriptions, \(r_t\) identifies the speaker of turn \(t\), \(x_t\) is the target text, and \(u_t\) specifies the associated local instructions.

Given \(\mathcal{S}\) and optional speaker-reference audio \(\mathcal{A}_{\mathrm{ref}}\) when supported, a system generates a scene-level audio output:
\[
Y \sim p_{\theta}
\left(
Y \mid \mathcal{S},\mathcal{A}_{\mathrm{ref}}
\right).
\]
The output should realize the scripted content and speaker assignments.

\subsection{DRAMA-Eval: Methods}
The evaluation target is the complete scene audio, whether produced through native dialogue generation or assembled from utterance-level outputs. This allows DRAMA-Eval to assess both local synthesis quality and discontinuities introduced across turns. We organize evaluation into three levels: objective measures of textual fidelity and speaker consistency; local judgments of instruction following, prosody, and expressive delivery; and global judgments of interaction naturalness, overall fluency, and scene immersion.

Following the pairwise evaluation paradigm, we compare model-generated speech with the corresponding real-world reference audio under the same evaluation criterion. The two audio samples are presented as A and B without revealing their sources, together with the relevant textual and contextual information. Rather than assigning an absolute score, the judge determines which sample better satisfies the target criterion or declares a tie. We aggregate these decisions into a tie-adjusted win rate, following the general pairwise evaluation principle used by EmergentTTS-Eval~\cite{emergent}, while using real-world audio as the reference. For a candidate system \(s\), we define
\[
\mathrm{WinRate}(s)
=
\frac{N_{\mathrm{win}}+0.5N_{\mathrm{tie}}}
{N_{\mathrm{comparisons}}},
\]
where \(N_{\mathrm{win}}\) counts comparisons in which the model-generated audio is preferred over the corresponding real-world audio, \(N_{\mathrm{tie}}\) counts ties, and \(N_{\mathrm{comparisons}}\) is the total number of valid comparisons. A win rate of \(0.5\) indicates parity with the real-world reference under the specified criterion and judging protocol; higher or lower values indicate a preference for generated or real-world audio, respectively. We report win rates separately for each evaluation dimension to distinguish instruction following, expressive delivery, and interaction-level quality.

To assess the reliability of Gemini 3.8 Flash~\cite{gemini38flash} as our automated judge, we validate its judgments against human subjective preferences. We also evaluate each audio pair twice with the A and B order reversed, counting a system as the winner only when its output is preferred in both presentations. This consistency requirement mitigates position bias and reduces sensitivity to presentation order. Furthermore, we average the resulting win rates across multiple independent evaluation runs to reduce variability in the judge's responses. Together, human agreement validation, order-swapped comparisons, and repeated evaluation provide a more reliable basis for system comparison.

\subsection{DRAMA-Eval: Dataset}

\paragraph{Diverse real-world sources.}
DRAMA-Eval is constructed from Chinese and English audio collected from diverse real-world sources, primarily podcasts and films or television series. Additional sources include interviews, news broadcasts, everyday conversations, livestream commerce, audiobooks, and advertisements. The film and television material spans genres such as historical drama, espionage, contemporary drama, and suspense. These sources cover both conversational and performed speech, providing varied character roles, expressive styles, and interactional contexts.

\paragraph{Segment-level audio annotation.}
We first use Gemini 2.5 Pro to annotate speaker-related attributes and local delivery characteristics for each speech segment. These annotations describe the voices and expressive behaviors observed in the source audio and provide the basis for constructing the evaluation scripts. This approach connects script-level descriptions and delivery requirements to real-world speech performances.

\paragraph{Structured script construction.}
We subsequently use DeepSeek-v4.1-Flash~\cite{deepseek-v4-1-flash} to clean the Gemini annotations and rewrite them into a consistent script format. The resulting scripts organize the information at three levels: global scene context, character profiles, and local delivery instructions associated with individual utterances. This organization provides generation systems with both the broader context of a scene and the specific expressive requirements of each turn.

\paragraph{Human quality control.}
Human reviewers inspect the resulting scripts for format compliance and the quality and suitability of their descriptions. The reviewed scripts form the final test inputs, while the corresponding real-world audio serves as the reference for evaluation. Figure~\ref{fig:drama_dataset_pipeline} summarizes the construction process.

\paragraph{Dataset statistics.}

DRAMA-Eval contains 1,038 cases, comprising 537 Chinese and 501 English cases. The source audio totals 12.50 hours, including 10.65 hours of annotated speech across 7,530 utterance-level turns. Each case lasts 43.36 seconds and contains 7.25 turns on average, while individual turns average 5.09 seconds. Cases contain one to four speakers and between 2 and 20 turns. Multi-speaker cases account for 65.6\% of the dataset, with two-speaker cases forming the largest group (478 cases). Most cases (68.0\%) contain 6--10 turns, enabling evaluation across successive utterances rather than isolated sentences. The dataset covers four evaluation categories: Normal (528 cases), Escalation (149), Contrast (162), and Rich Emotions (199). Turn-level emotion annotations include calm speech, sadness, happiness, anger, fear, surprise, and disgust. Figure~\ref{fig:dataset_dist} summarizes the distributions of evaluation categories, speaker counts, turn counts, and emotions.

\section{Experiments}

\begin{table*}[t]
\centering
\caption{
DRAMA-Eval results on Chinese.
\textbf{Objective metrics:}
A1 measures textual fidelity using character error rate (CER, \%; lower is better);
A2 measures speaker consistency across utterances (higher is better).
\textbf{Local evaluation:}
B1 assesses instruction following,
B2 assesses utterance-level prosody and fluency,
and B3 assesses local expressive delivery.
\textbf{Global evaluation:}
C1 assesses interaction naturalness,
C2 assesses overall fluency across turns,
and C3 assesses scene immersion.
B1--C3 are LALM-evaluated, tie-adjusted win rates against the corresponding real-world audio, reported in percentage points with the \% sign omitted; 50 indicates parity with the reference under the evaluation protocol.
These scores are averaged over three independent evaluation runs, with accompanying $\pm$ values indicating run-to-run variability.
Dashes indicate that instruction-following evaluation is not reported for models without an instruction-control interface.
Best mean results are highlighted in bold.
}
\label{tab:drama_zh}
\begingroup
\small
\setlength{\tabcolsep}{3pt}
\renewcommand{\arraystretch}{1.12}
\resizebox{\textwidth}{!}{%
\begin{tabular}{lcccccccc}
\toprule
Model & A1$\downarrow$ & A2$\uparrow$
& B1$\uparrow$ & B2$\uparrow$ & B3$\uparrow$
& C1$\uparrow$ & C2$\uparrow$ & C3$\uparrow$ \\
\midrule

\rowcolor{green!15}
SeedAudio~\cite{seedaudio2024}
& \textbf{2.58} & $0.74\pm0.05$
& $\mathbf{62.2}\pm0.4$ & $57.4\pm0.2$
& $\mathbf{64.2}\pm0.6$ & $\mathbf{67.4}\pm0.7$
& $\mathbf{73.2}\pm0.4$ & $\mathbf{75.7}\pm1.1$ \\

\rowcolor{green!15}
StepAudio 3 Gen~\cite{step-audio3-gen}
& 3.78 & $0.76\pm0.05$
& $57.3\pm0.6$ & $\mathbf{58.3}\pm0.3$
& $56.9\pm0.2$ & $60.2\pm1.7$
& $64.7\pm1.5$ & $64.6\pm1.4$ \\

\rowcolor{blue!15}
MiniMax-H3~\cite{minimaxh3_2025}
& 3.84 & $0.58\pm0.12$
& $56.6\pm0.2$ & $57.9\pm0.2$
& $57.6\pm0.3$ & $49.7\pm0.1$
& $42.2\pm0.8$ & $43.2\pm0.6$ \\

\rowcolor{blue!15}
Qwen3-TTS (VD)~\cite{qwentts}
& 2.73 & $0.59\pm0.08$
& $59.7\pm0.2$ & $54.6\pm0.5$
& $60.6\pm0.5$ & $40.0\pm0.0$
& $28.2\pm0.0$ & $27.7\pm0.0$ \\

\rowcolor{blue!15}
FireRedTTS3~\cite{fireredtts3}
& 2.64 & $0.62\pm0.07$
& $52.1\pm0.3$ & $55.1\pm0.4$
& $50.4\pm0.7$ & $39.8\pm0.7$
& $31.9\pm0.8$ & $30.7\pm1.4$ \\

\rowcolor{yellow!15}
VoxCPM2~\cite{voxcpm2}
& 3.25 & $0.67\pm0.06$
& $48.1\pm0.5$ & $48.6\pm0.3$
& $48.0\pm0.7$ & $36.4\pm0.2$
& $29.8\pm0.5$ & $29.1\pm0.8$ \\

\rowcolor{yellow!15}
MiMo-Audio-7B~\cite{mimo-audio}
& 5.17 & $0.55\pm0.08$
& $36.8\pm1.1$ & $34.8\pm1.2$
& $37.0\pm1.3$ & $24.2\pm0.2$
& $10.8\pm0.2$ & $10.9\pm0.1$ \\

\rowcolor{red!15}
MOSS-TTSD~\cite{moss-ttsd}
& 4.06 & $\mathbf{0.78}\pm0.04$
& $--$ & $46.6\pm0.2$
& $35.9\pm1.1$ & $38.1\pm0.5$
& $38.8\pm0.2$ & $35.2\pm0.7$ \\

\rowcolor{red!15}
SoulX-Podcast~\cite{soulxpodcast}
& 2.74 & $0.78\pm0.04$
& $--$ & $40.7\pm0.5$
& $27.1\pm0.4$ & $35.4\pm0.6$
& $31.3\pm0.6$ & $27.7\pm0.4$ \\

\rowcolor{red!15}
VibeVoice~\cite{vibevoice}
& 5.62 & $0.68\pm0.07$
& $--$ & $35.0\pm0.6$
& $31.8\pm0.2$ & $28.7\pm0.9$
& $22.4\pm0.5$ & $21.5\pm0.6$ \\

\rowcolor{red!15}
FireRedTTS2~\cite{fireredtts2}
& 3.01 & $0.73\pm0.05$
& $--$ & $39.5\pm0.6$
& $28.8\pm0.3$ & $31.7\pm1.0$
& $27.4\pm1.3$ & $25.8\pm1.0$ \\

\rowcolor{red!15}
Higgs Audio V2~\cite{higgsaudio2025}
& 8.43 & $0.70\pm0.07$
& $--$ & $24.1\pm0.1$
& $19.8\pm0.4$ & $23.7\pm0.2$
& $13.4\pm0.6$ & $12.4\pm0.2$ \\

\bottomrule
\end{tabular}
}
\endgroup
\end{table*}

\begin{table*}[t]
\centering
\caption{
DRAMA-Eval results on English.
A1 reports WER (\%); all other metrics and reporting conventions follow Table~\ref{tab:drama_zh}.
Best mean results are highlighted in bold.
}
\label{tab:drama_en}
\begingroup
\small
\setlength{\tabcolsep}{3pt}
\renewcommand{\arraystretch}{1.12}
\resizebox{\textwidth}{!}{%
\begin{tabular}{lcccccccc}
\toprule
Model & A1$\downarrow$ & A2$\uparrow$
& B1$\uparrow$ & B2$\uparrow$ & B3$\uparrow$
& C1$\uparrow$ & C2$\uparrow$ & C3$\uparrow$ \\
\midrule

\rowcolor{green!15}
SeedAudio~\cite{seedaudio2024}
& 4.45 & $0.68\pm0.07$
& $\mathbf{56.5}\pm0.8$ & $\mathbf{50.5}\pm0.2$
& $\mathbf{57.0}\pm0.7$ & $\mathbf{56.8}\pm1.0$
& $\mathbf{54.5}\pm0.4$ & $\mathbf{55.7}\pm0.8$ \\

\rowcolor{green!15}
StepAudio 3 Gen~\cite{step-audio3-gen}
& 6.00 & $0.67\pm0.10$
& $48.5\pm0.4$ & $48.1\pm0.4$
& $45.2\pm0.2$ & $43.4\pm0.5$
& $33.8\pm0.6$ & $33.6\pm0.5$ \\

\rowcolor{blue!15}
MiniMax-H3~\cite{minimaxh3_2025}
& 10.56 & $0.44\pm0.16$
& $50.9\pm0.3$ & $\mathbf{50.5}\pm0.2$
& $51.6\pm0.6$ & $37.8\pm0.3$
& $22.4\pm0.6$ & $23.9\pm0.9$ \\

\rowcolor{blue!15}
Qwen3-TTS (VD)~\cite{qwentts}
& 4.16 & $0.40\pm0.10$
& $46.9\pm0.3$ & $48.3\pm0.3$
& $44.5\pm0.3$ & $34.3\pm0.0$
& $16.1\pm0.0$ & $15.0\pm0.0$ \\

\rowcolor{blue!15}
FireRedTTS3~\cite{fireredtts3}
& \textbf{3.97} & $0.47\pm0.09$
& $37.0\pm0.2$ & $46.2\pm0.4$
& $30.7\pm0.3$ & $30.9\pm1.1$
& $13.2\pm0.6$ & $11.5\pm0.5$ \\

\rowcolor{yellow!15}
VoxCPM2~\cite{voxcpm2}
& 8.77 & $0.69\pm0.06$
& $32.5\pm0.4$ & $37.9\pm0.0$
& $28.5\pm0.7$ & $28.9\pm0.6$
& $16.7\pm0.4$ & $14.8\pm0.6$ \\

\rowcolor{yellow!15}
MiMo-Audio-7B~\cite{mimo-audio}
& 9.12 & $0.45\pm0.09$
& $28.3\pm0.3$ & $28.5\pm0.2$
& $25.5\pm0.1$ & $24.6\pm0.3$
& $5.9\pm0.5$ & $6.3\pm0.7$ \\

\rowcolor{red!15}
MOSS-TTSD~\cite{moss-ttsd}
& 6.03 & $\mathbf{0.77}\pm0.04$
& $--$ & $40.4\pm0.3$
& $23.5\pm0.2$ & $31.4\pm0.3$
& $25.3\pm0.8$ & $18.9\pm0.7$ \\

\rowcolor{red!15}
SoulX-Podcast~\cite{soulxpodcast}
& 4.50 & $0.76\pm0.05$
& $--$ & $32.1\pm0.4$
& $16.7\pm0.1$ & $27.7\pm0.3$
& $16.2\pm0.3$ & $12.8\pm0.4$ \\

\rowcolor{red!15}
VibeVoice~\cite{vibevoice}
& 9.81 & $0.64\pm0.08$
& $--$ & $32.8\pm0.2$
& $23.2\pm0.3$ & $27.1\pm0.6$
& $15.4\pm0.5$ & $13.3\pm0.5$ \\

\rowcolor{red!15}
FireRedTTS2~\cite{fireredtts2}
& 6.12 & $0.66\pm0.07$
& $--$ & $29.0\pm0.6$
& $16.2\pm0.5$ & $25.8\pm0.4$
& $10.6\pm0.2$ & $8.7\pm0.4$ \\

\rowcolor{red!15}
Higgs Audio V2~\cite{higgsaudio2025}
& 8.05 & $0.69\pm0.07$
& $--$ & $28.9\pm0.3$
& $19.6\pm0.1$ & $26.4\pm0.3$
& $15.0\pm0.7$ & $12.6\pm1.0$ \\

\bottomrule
\end{tabular}
}
\endgroup
\end{table*}

\label{sec:experiments}

\subsection{Inference Protocols}
\label{sec:inference}

We adopt capability-specific inference protocols for four categories of speech generation systems. All systems synthesize the target utterance texts, while additional conditioning inputs depend on their supported interfaces.

\paragraph{Dialogue TTS.}
These systems receive the target texts, speaker assignments, and reference audio, without global scene descriptions or local delivery instructions. Reference utterances are sampled from Emilia~\cite{emilia} to specify the target voices. This configuration evaluates voice maintenance during dialogue generation and reveals how expressive and interactional performance develops without explicit contextual guidance.

\paragraph{Single-speaker voice design.}
These systems generate speech without reference audio. For each utterance, we concatenate the corresponding global character description and local style description into a natural-language instruction. Utterances are synthesized individually and concatenated in script order to form the complete scene.

\paragraph{Single-speaker voice control.}
These systems receive both speaker-reference audio and the corresponding delivery instructions. Each utterance is generated separately, and the resulting utterances are concatenated in script order. This configuration evaluates whether reference-conditioned voice control can maintain speaker consistency while realizing utterance-level expressive requirements.

\paragraph{Context-aware multi-speaker generation.}
These systems receive the complete structured script, including global scene context, character descriptions, speaker assignments, and local delivery instructions, and generate the scene as a whole. This configuration allows the model to condition expressive choices and turn transitions on the surrounding interaction.

These protocols evaluate each system through its supported conditioning interface. Cross-category comparisons therefore characterize the capabilities of the evaluated generation configurations, rather than performance under identical contextual inputs.

The evaluated models are organized into four color-coded groups, ordered from top to bottom in the tables.
\textbf{Green} rows denote context-aware multi-speaker generation models, which receive the complete structured script and generate the scene as a whole.
\textbf{Blue} rows denote single-speaker voice-design models, which generate individual utterances from instructions combining global character descriptions and local style descriptions, without reference audio.
\textbf{Yellow} rows denote single-speaker voice-control models, which receive both speaker-reference audio and delivery instructions.
Outputs from both single-speaker categories are concatenated in script order.
\textbf{Red} rows denote multi-speaker TTS models, which receive target texts, speaker assignments, and reference audio, without global scene context or local delivery instructions.
These groups distinguish the conditioning interfaces and generation protocols used in our evaluation.
\subsection{Objective Evaluation}
\label{sec:objective_eval}

We evaluate textual fidelity and speaker consistency using two objective metrics. A1 reports word error rate (WER) for English and character error rate (CER) for Chinese, computed using Qwen3-0.6B-ASR~\cite{qwen3asr}. Lower values indicate better preservation of the target text. A2 measures speaker consistency across utterances assigned to the same character using speaker embeddings following the Seed-TTS-Eval protocol~\cite{seed-tts-eval}. Unlike similarity to a reference recording, this metric assesses whether a character maintains a consistent voice throughout the generated scene.

SeedAudio achieves the lowest Chinese CER of 2.58, while FireRedTTS3 achieves the lowest English WER of 3.97. Strong textual fidelity is observed across different generation categories, suggesting that accurate transcription alone does not distinguish systems capable of expressive scene generation.

Dialogue TTS systems exhibit strong speaker consistency. MOSS-TTSD obtains approximately 0.78 on Chinese and 0.77 on English, while SoulX-Podcast reaches approximately 0.78 and 0.76, respectively. However, their strong identity consistency does not translate into equally strong expressive and interaction-level results. Conversely, several voice-design configurations achieve competitive textual fidelity but lower speaker consistency across independently generated utterances. These results motivate evaluating identity maintenance separately from both content accuracy and expressive quality.

\subsection{LALM-Based Evaluation}
\label{sec:lalm_eval}

We use Gemini 3.8 Flash as an audio-language judge to evaluate six dimensions: instruction following (B1), prosody and fluency (B2), local expressiveness (B3), interaction naturalness (C1), overall fluency (C2), and scene immersion (C3). For each dimension, the judge compares model-generated audio with the corresponding real-world reference audio, using the relevant script-level information.

Each audio pair is evaluated twice with the A/B order reversed. A system is counted as the winner only when its output is preferred in both presentations. We report tie-adjusted win rates, assigning half credit to ties, and average the results over three independent evaluation runs. Scores are reported in percentage points; 50 indicates parity with the real-world reference under the specified criterion and judging protocol.

\paragraph{Local expressive performance.}
SeedAudio achieves the highest instruction-following and local-expressiveness scores in both languages, reaching 62.2/56.5 on B1 and 64.2/57.0 on B3 for Chinese/English. StepAudio 3 Gen achieves the highest Chinese B2 score of 58.3, indicating that the strongest system varies across expressive dimensions. Voice-design models also demonstrate competitive local control: Qwen3-TTS reaches 59.7 on Chinese B1 and 60.6 on Chinese B3.

\paragraph{Scene-level interaction quality.}
The differences become more pronounced at the scene level. SeedAudio achieves C1/C2/C3 scores of 67.4/73.2/75.7 on Chinese and 56.8/54.5/55.7 on English. In contrast, Qwen3-TTS obtains Chinese C2/C3 scores of 28.2/27.7 despite its competitive local expressive scores. This gap illustrates that successful utterance-level instruction execution does not necessarily produce a coherent scene after concatenation.

Dialogue TTS systems show a complementary limitation. MOSS-TTSD maintains strong speaker consistency but obtains Chinese B3 scores of 35.9 under its reference-only configuration. Together, these results reveal distinct capability gaps: voice maintenance does not ensure expressive delivery, and local controllability does not ensure scene-level continuity. Context-aware generation achieves stronger overall scene performance in the evaluated configurations, highlighting the value of assessing local and global qualities jointly.

\subsection{Agreement with Human Judgments}
\label{sec:human_agreement}

We assess the automated judge against human subjective preferences using three labels: generated audio preferred, tie, and real-world audio preferred. Agreement receives a score of 1 when the human and judge labels match, 0.5 when one selects a tie and the other expresses a preference, and 0 when their preferences are opposite.

We report two aggregation methods. Annotator-level agreement compares each individual human annotation with the judge and averages the resulting scores. Consensus-level agreement first selects the most frequent human label for each case and dimension, retaining it only when it receives at least two votes, and then compares it with the judge. Table~\ref{tab:human_llm_agreement} reports both results.

Annotator-level agreement ranges from 65.23\% to 73.05\%, while consensus-level agreement ranges from 65.91\% to 81.58\%. Agreement is higher with consensus labels across all six dimensions, with the highest value observed for interaction naturalness (C1). These results support the automated judge as a useful basis for relative system comparison, while the remaining disagreement motivates reporting its scores together with human validation.

\begin{table}[t]
\centering
\small
\setlength{\tabcolsep}{5pt}
\caption{
Human--LLM partial-credit agreement (\%) for the judge.
Exact matches score 1, tie/non-tie mismatches score 0.5,
and opposite preferences score 0.
Parentheses report the number of annotation-level comparisons
for Individual and valid cases for Consensus.
Consensus uses the most frequent human judgment, requiring
at least two matching votes.
}
\label{tab:human_llm_agreement}
\begin{tabular}{lcc}
\toprule
Dimension & Individual & Consensus \\
\midrule
B1 & 65.23 (256) & 65.91 (66) \\
B2 & 66.12 (256) & 67.53 (65) \\
B3 & 66.02 (256) & 71.85 (69) \\
\midrule
C1 & 73.05 (154) & 81.58 (38) \\
C2 & 69.27 (192) & 77.27 (44) \\
C3 & 69.27 (192) & 76.19 (42) \\
\bottomrule
\end{tabular}
\end{table}

\subsection{Global Context Ablation}
We ablate global context on SeedAudio and evaluate the resulting
scene-level quality using AB comparisons against real-world audio.
As shown in Table~\ref{tab:global_context_ablation}, providing global
context improves all three dimensions in both languages. These results support the value of global context for scene-level generation, particularly for producing speech perceived as appropriate to the overall scene.
\begin{table}[t]
\centering
\small
\setlength{\tabcolsep}{5pt}
\caption{
Ablation of global context on SeedAudio. C1--C3 are reported.
}
\label{tab:global_context_ablation}
\begin{tabular}{llccc}
\toprule
Lang. & Global context & C1$\uparrow$ & C2$\uparrow$ & C3$\uparrow$ \\
\midrule

ZH & Without & 61.1 & 66.2 & 64.4 \\
   & With    & \textbf{67.4} & \textbf{73.2} & \textbf{75.7} \\
\midrule
EN & Without & 50.8 & 48.5 & 45.9 \\
   & With    & \textbf{56.8} & \textbf{54.5} & \textbf{55.7} \\
\bottomrule
\end{tabular}
\end{table}

\section{Conclusion}
We introduced DRAMA-Eval, a bilingual benchmark for evaluating context-aware, multi-speaker speech generation from structured scripts grounded in real-world audio. Its evaluation framework jointly assesses textual fidelity, speaker consistency, local instruction execution, and scene-level interaction quality. Experiments across four generation paradigms reveal complementary limitations in expressive control, voice consistency, and coherent interaction. Human agreement analysis supports the automated judging protocol, while context ablations highlight the importance of global scene information. DRAMA-Eval provides a unified testbed for advancing controllable and coherent speech generation.

\section{Limitations}
DRAMA-Eval currently covers only Chinese and English, limiting generalization to other languages and cultural contexts. Its automated evaluation relies on a proprietary audio-language model, whose biases and updates may affect reliability and reproducibility despite human validation. The dataset is also imbalanced across emotions and speaker configurations, and its scene lengths do not capture extended narrative generation. The current evaluation focuses on speech generation and does not assess complex audio generation involving background ambience, music, or sound effects, including their quality and coordination with speech. Finally, models receive different conditioning inputs according to their supported interfaces, which should be considered when interpreting cross-category comparisons.

\clearpage
\appendix
\section{Evaluation Prompts and Examples}
\subsection{Local LALM Evaluation Prompt}
\label{app:local_judging_prompts}

The following prompt evaluates instruction following (B1),
utterance-level prosody and fluency (B2), and local
expressiveness (B3) through pairwise audio comparisons.
Each dimension is judged independently, with a preference
for Audio A, Audio B, or a tie.

\begin{tcolorbox}[
    enhanced,breakable,title={Local Evaluation: B1--B3},
    colback=black!2,colframe=black!35,
    colbacktitle=black!8,coltitle=black,
    fonttitle=\small\bfseries,boxrule=0.4pt,
    left=5pt,right=5pt,top=5pt,bottom=5pt]
\begin{CJK*}{UTF8}{gbsn}
\begin{Verbatim}[
    fontfamily=cmtt,
    fontsize=\footnotesize,breaklines=true,
    breakanywhere=false,breaksymbolleft={},frame=none
]
<role>
You are an elite, highly objective audio drama director and acoustic analyst. Your task is to perform a blind A/B test between two synthesized audio clips (Audio A and Audio B) rendering the exact same script turn by turn.
</role>

<workflow>
For EVERY turn in the `content` list provided in the `<script_data>`:
1. [Planner]: Read the character's `text_info` and the current turn's `context` (stage directions) to deduce the EXPECTED acoustic delivery (e.g., emotion, speed, breath, tone).
2. [Judge]: Listen to Audio A and Audio B for that specific turn. 
3. [Decoupled Comparison]: Compare the two audios across three strictly decoupled local metrics (B1, B2, B3).
4. [Declare Winner]: Output "A", "B", or "Tie".
</workflow>

<critical_rules>
**RULE 1: THE BALANCED "TIE" THRESHOLD**
- **Do not force a winner for trivialities**: If both audios are equally generic TTS readings, or both flawlessly execute the instruction with only negligible volume differences, you MUST output "Tie".
- **Do not abuse "Tie" out of caution**: If one audio clearly captures the instructed emotion better, has fewer robotic glitches, or demonstrates significantly deeper acting tension, you MUST decisively reward it.

**RULE 2: STRICT DECOUPLING OF METRICS**
- B1 ONLY judges *Fidelity to Instruction* (Did it do what the `context` asked?).
- B2 ONLY judges *Physical Acoustic Quality* (Are there robotic glitches, metallic sounds, or unnatural stutters?).
- B3 ONLY judges *Acting Depth and Dynamic Range* (Does it sound like a professional actor or a flat machine?).
*Note: An audio might completely fail B1 (wrong emotion) but win B3 (very dynamic and dramatic, just the wrong vibe).*

**RULE 3: TEXT SEMANTICS vs. STAGE DIRECTIONS**
- If the semantic meaning of the spoken `text` conflicts with the `context` (e.g., text says "I hate you" but context says "He says it playfully and laughing"), you MUST judge based on the `context`.

**RULE 4: BGM & NOISE IMMUNITY**
- Ground Truth audio may contain background music (BGM) or room noise. AI audio might be completely dry.
- Do NOT penalize B2 for natural environmental noise. B2 ONLY penalizes AI-generated mechanical glitches.
- Do NOT reward B1/B3 just because an audio has dramatic BGM. Judge the HUMAN VOCAL CORDS ONLY.
</critical_rules>

<metrics_definitions>
**Metric B1: Instruction Following & Context Fidelity**
- **Concept**: How accurately does the voice physicalize the specific instructions in the `context` and `text_info`?
- **What to listen for**:
  - Focus on specific actions, emotions, and attitudes requested (e.g., whispering, shouting, crying, sighing, authoritative tone, rapid speech).
  - *Winner*: The audio that more precisely matches the requested intent. Ignore audio glitches for this metric.

**Metric B2: Prosody & Fluency**
- **Concept**: Which audio has a more natural human rhythm, phrasing, and zero mechanical synthesis artifacts?
- **What to listen for**:
  - Focus purely on the physical synthesis quality. 
  - Look for unnatural stutters, metallic buzzing, clipping, weird mid-word pauses, or robotic intonation.
  - *Winner*: The audio with the cleaner, more human-like physical signal and rhythm. Ignore whether the emotion is correct.

**Metric B3: Local Expressiveness**
- **Concept**: Which audio sounds more like a professional voice actor with deep artistic tension?
- **What to listen for**:
  - Focus on dynamic range (variations in pitch, energy, and pace) and micro-expressions (breaths, vocal fry).
  - Does it sound like a rich, theatrical performance or a monotone, generic TTS reading the news?
  - *Winner*: The audio that is less flat and exhibits more sophisticated vocal acting.
</metrics_definitions>

<output_format>
Return ONLY a valid JSON object. Do not use Markdown formatting (no ```json).
The "winner" field MUST be EXACTLY one of: "A", "B", or "Tie".
{
  "utterance_evaluations": [
    {
      "turn_num": 1,
      "B1_Instruction_Following": {
        "expected_delivery_analysis": "[Based on the `context`, what specific acoustic traits are expected?]",
        "audio_a_obs": "[Did A perform these traits?]",
        "audio_b_obs": "[Did B perform these traits?]",
        "comparison_reasoning": "[Who is more accurate to the instruction?]",
        "winner": "Tie"
      },
      "B2_Prosody_Fluency": {
        "audio_a_obs": "[Rhythm and AI artifacts in A]",
        "audio_b_obs": "[Rhythm and AI artifacts in B]",
        "comparison_reasoning": "[Who has cleaner, more natural physical synthesis?]",
        "winner": "A"
      },
      "B3_Local_Expressiveness": {
        "audio_a_obs": "[Dynamic range and acting depth in A]",
        "audio_b_obs": "[Dynamic range and acting depth in B]",
        "comparison_reasoning": "[Who sounds more like a professional actor rather than a flat reader?]",
        "winner": "B"
      }
    }
  ]
}
</output_format>
\end{Verbatim}
\end{CJK*}
\end{tcolorbox}

\subsection{Global LALM Evaluation Prompt}
\label{app:global_judging_prompts}

The following prompt evaluates interaction naturalness (C1),
overall fluency (C2), and scene immersion (C3) through
pairwise comparisons of complete scene audio.
The structured script and audio samples are supplied separately,
and each dimension receives an independent A/B/tie judgment.

\begin{tcolorbox}[
    enhanced,breakable,title={Global Evaluation: C1--C3},
    colback=black!2,colframe=black!35,
    colbacktitle=black!8,coltitle=black,
    fonttitle=\small\bfseries,boxrule=0.4pt,
    left=5pt,right=5pt,top=5pt,bottom=5pt]
\begin{CJK*}{UTF8}{gbsn}
\begin{Verbatim}[
    fontfamily=cmtt,
    fontsize=\footnotesize,breaklines=true,
    breakanywhere=false,breaksymbolleft={},frame=none
]
<role>
You are an elite, highly objective audio drama director and acoustic analyst. Your task is to perform a blind A/B test between two synthesized long-form audio clips (Audio A and Audio B) based on the provided script.
</role>

<workflow>
1. [Script Comprehension]: Carefully read the provided `<script_data>`. You must understand the `context_background` (overall scene), the `character` profiles (expected voice traits), and the turn-by-turn `content` (speaker identity, spoken text, and local stage directions).
2. [Macro Listening]: Listen to the entirety of Audio A and Audio B. Do not focus merely on isolated words; focus on the macro-level flow, how speakers interact, and the overall theatrical vibe.
3. [Decoupled Comparison]: Compare the two audios across three strictly decoupled global metrics (C1, C2, C3).
4. [Declare Winner]: Output "A", "B", or "Tie".
</workflow>

<critical_rules>
**RULE 1: THE BALANCED "TIE" THRESHOLD (CRITICAL)**
- **Do not force a winner for trivialities**: If the differences are microscopic (e.g., a slight volume variance) and both audios equally fail or succeed at the core metric, you MUST output "Tie".
- **Do not abuse "Tie" out of caution**: If one audio demonstrates a clear, discernible advantage in capturing the interaction, maintaining acoustic flow, or building the scene, you MUST courageously reward it. Do not default to "Tie" just to avoid making a hard choice. Be accurate, objective, and decisive.

**RULE 2: STRICT DECOUPLING OF METRICS**
- C1 ONLY judges the *pragmatic interaction and feedback* between DIFFERENT speakers.
- C2 ONLY judges the *organic, cohesive physical listening experience* (splicing, acoustic continuity).
- C3 ONLY judges the *semantic-driven theatrical immersion* (does it feel like a real audiobook/drama?).

**RULE 3: BGM & NOISE IMMUNITY**
- Ground Truth audio may contain background music (BGM) or room echo. AI audio might be completely dry.
- Do NOT penalize C2 for continuous natural environmental noise. 
- Do NOT reward C3 just because an audio has dramatic BGM. Judge the HUMAN VOCAL PERFORMANCE, TIMING, AND DELIVERY ONLY.
</critical_rules>

<metrics_definitions>
**Metric C1: Cross-speaker Interaction Naturalness**
- **Concept**: By looking at the speaker identities in the script, do the different characters sound like they are actually communicating with each other in the same spatial and conversational context?
- **What to listen for**:
  - *Contextual Feedback*: Does the responding speaker sound like they *heard* the previous speaker? 
  - *Agreement/Continuation*: If agreeing or echoing, does the tone naturally build upon the previous speaker's energy?
  - *Conflict/Rebuttal*: If arguing or shifting topics, does the interruption, contrast in tone, or escalation feel logically driven by the previous line?
  - *Bad Interaction*: Sounds like two isolated people reading lines in different rooms, completely ignoring each other's semantic and acoustic cues (e.g., disjointed pragmatics, front-to-back incoherence).
- **Constraint**: If there is only ONE speaker in the entire script, this metric is NOT APPLICABLE. You must state "Single speaker, skipped" in the analysis and force the winner to "Tie".

**Metric C2: Global Flow & Acoustic Continuity**
- **Concept**: Subjectively, does the entire track feel organic and cohesive, as if recorded in a single continuous take or a natural conversation?
- **What to listen for**:
  - Do not focus on semantics or expressiveness here. Focus purely on the macro auditory flow.
  - *Bad Flow*: Sentences feel mechanically glued together. Noticeable splicing artifacts, abrupt cuts in background room-tone (dead air), or sudden jarring shifts in audio quality/EQ between sentences.
  - *Good Flow*: The transitions between sentences and breaths feel natural, integrated, and completely seamless.

**Metric C3: Scene Immersion & Theatrical Fidelity**
- **Concept**: Does the collective semantic delivery and acting energy successfully transport the listener into the `context_background`?
- **What to listen for**:
  - Just as pitch, speed, and volume are cues for emotion, the acoustic continuity, emotional trajectory, and character consistencies are merely underlying factors that contribute to the overall immersion. 
  - Evaluate the holistic vibe: Does it feel like a professional audiobook or a high-quality radio drama?
  - Are all the utterances cohesive in establishing the *same* overarching scene? (e.g., If the scene is a tense police meeting, does the entire audio maintain that high-stakes, suppressed energy based on the script?)
  - *Bad Immersion*: Characters randomly switch emotions without narrative logic, failing to establish the stated atmosphere.
</metrics_definitions>

<output_format>
Return ONLY a valid JSON object. Do not use Markdown formatting (no ```json).
The "winner" field MUST be EXACTLY one of: "A", "B", or "Tie".
{
  "C1_Interaction_Naturalness": {
    "interaction_analysis": "[Analyze if speakers sound like they are reacting to each other (agreement/conflict/feedback) based on the script. If single speaker, write 'Single speaker, skipped']",
    "audio_a_obs": "[Observations on A's pragmatic interaction]",
    "audio_b_obs": "[Observations on B's pragmatic interaction]",
    "comparison_reasoning": "[Who has a more logically coherent and natural conversational feedback?]",
    "winner": "Tie"
  },
  "C2_Global_Flow": {
    "cohesion_analysis": "[Analyze the organic wholeness vs. mechanical splicing of the track]",
    "audio_a_obs": "[Observations on A's physical flow and continuity]",
    "audio_b_obs": "[Observations on B's physical flow and continuity]",
    "comparison_reasoning": "[Who sounds more like a cohesive, natural single take?]",
    "winner": "A"
  },
  "C3_Scene_Immersion": {
    "theatrical_fidelity_analysis": "[Analyze how well the combined semantics and acting build the `context_background` and script's trajectory]",
    "audio_a_obs": "[Observations on A's audiobook/drama immersion]",
    "audio_b_obs": "[Observations on B's audiobook/drama immersion]",
    "comparison_reasoning": "[Who better establishes the overarching scene and holistic theatrical vibe?]",
    "winner": "B"
  }
}
</output_format>
\end{Verbatim}
\end{CJK*}
\end{tcolorbox}

\subsection{Script Example}

\begin{tcolorbox}[
    enhanced,breakable,title={Script Example},
    colback=black!2,colframe=black!35,
    colbacktitle=black!8,coltitle=black,
    fonttitle=\small\bfseries,boxrule=0.4pt,
    left=5pt,right=5pt,top=5pt,bottom=5pt]
\begin{CJK*}{UTF8}{gbsn}
\begin{Verbatim}[
    fontfamily=cmtt,
    fontsize=\footnotesize,breaklines=true,
    breakanywhere=false,breaksymbolleft={},frame=none
]
{
  "utt": "0008",
  "language": "中文",
  "audio_path": "",
  "eval_type": "Contrast",
  "context_meta": {
    "total_turns": 4,
    "total_chars": 100
  },
  "context_background": "警局会议室内，白炽灯下烟雾缭绕，几名警务人员围坐在长桌旁，墙上挂着城市地图。气氛凝重，众人正分析一起解款车劫案。",
  "character": {
    "汇报男_中年": {
      "text_info": "中年男性，声音洪亮，语速偏快，吐字清晰，带有职业性的汇报腔调。",
      "wav_path": "",
      "prompt_slot": "Middle_Male_Neutral"
    },
    "警官男_中年": {
      "text_info": "中年男性，声音坚实低沉，富有权威感，语气果断，带有不容置疑的指令性。",
      "wav_path": "",
      "prompt_slot": "Middle_Male_Happy"
    }
  },
  "content": [
    {
      "turn_num": 1,
      "text": "解款车的路线图会放在这个信封里面，出发前交给负责押运的经理，他会在上车后才打开信封，也就是说，事前没有人知道行车路线。",
      "speaker_info": {
        "speaker_name": "汇报男_中年",
        "gender": "男声",
        "age": "中年"
      },
      "audio_assets": {
        "start_time_ms": 0,
        "end_time_ms": 8480,
        "duration_ms": 8480
      },
      "context": "他清了清嗓子，目光扫过众人，用洪亮的声音快速而清晰地说明流程，试图强调安保的严密性。"
    },
    {
      "turn_num": 2,
      "text": "车没有按指定路线，一定是车上有内鬼。",
      "speaker_info": {
        "speaker_name": "警官男_中年",
        "gender": "男声",
        "age": "中年"
      },
      "audio_assets": {
        "start_time_ms": 11600,
        "end_time_ms": 14880,
        "duration_ms": 3280
      },
      "context": "他眉头紧锁，双手撑在桌沿，身体前倾，用不容置疑的语气果断下结论，目光锐利地扫视众人。"
    {
      "turn_num": 3,
      "text": "大家记住，死的人只是线。",
      "speaker_info": {
        "speaker_name": "警官男_中年",
        "gender": "男声",
        "age": "中年"
      },
      "audio_assets": {
        "start_time_ms": 17120,
        "end_time_ms": 20280,
        "duration_ms": 3160
      },
      "context": "他压低声音，眼神变得锐利，带着警告意味，用沉稳有力的语气强调关键，仿佛在敲打每个人的神经。"
    },
    {
      "turn_num": 4,
      "text": "现在最重要的是找到他。",
      "speaker_info": {
        "speaker_name": "警官男_中年",
        "gender": "男声",
        "age": "中年"
      },
      "audio_assets": {
        "start_time_ms": 23020,
        "end_time_ms": 24740,
        "duration_ms": 1720
      },
      "context": "他直起身，目光坚定地看向众人，用目标明确的指令语气布置任务，语气中带着不容拖延的紧迫感。"
    }
  ]
}
\end{Verbatim}
\end{CJK*}
\end{tcolorbox}

\end{document}